\documentclass{PoS}
\usepackage{graphicx}
\usepackage{amsmath}%

\usepackage{amsfonts}%
\usepackage{amssymb}%

\PoS{PoS(Lattice 2005)153}
 
\FullConference{$23^{rd}$ International Symposium on Lattice Field Theory\\
Trinity college, Dublin\\
July 23-29, 2005}
\author{ \speaker{K.~Petrov},\\
Physics Department, Brookhaven National Laboratory, Upton, NY, 11973,  USA\\
E-mail: \email{petrov@bnl.gov}}
\author{A.~Jakov\'ac\\
Institute of Physics, BME Budapest, Budafoki ú\'ut 8, H-1111 Budapest,
Hungary\\
\email{Antal.Jakovac@cern.ch}}
\author{P.~Petreczky\\
Physics Department, Brookhaven National Laboratory, Upton, NY, 11973,  USA\\
\email{petreczk@bnl.gov}}
\author{A.~Velytsky\\
Department of Physics and Astronomy, UCLA, Los Angeles, CA 90095-1547, USA\\
\email{vel@physics.ucla.edu}}

\title{Bottomonia correlators and spectral functions at zero and finite 
temperature}
\ShortTitle{Quarkonium Spectral Functions at zero and finite temperature}

\abstract{
We present 
preliminary studies of bottomonia spectral functions at
zero and finite temperature using quenched anisotropic
lattices. The heavy quark is treated within Fermilab
approach. We find no modification of the $\eta_b$ and
$\Upsilon$ states up to temperatures $2.3T_c$ while 
our study suggest dissolution of $\chi_b$ state 
at $1.15T_c$.
} 

\begin{document}

\section{Introduction}
\label{intro}
Long ago it was conjectured by Matsui and Satz \cite{MS86} that melting of
different quarkonia states due to color screening 
can signal Quark Gluon Plasma formation in heavy ion collisions.
The problem of quarkonia dissolution has been studied using potential models
\cite{karsch88,digal01a,digal01b,shuryak04,wong04,mocsyhard04}.
However it is very unclear if such models are valid 
at the finite temperature  \cite{petreczkyhard04}. Therefore the problem of
dissolution of different quarkonia states should be studied in terms of
the corresponding meson spectral functions. For charmonium such studies
appeared only recently and suggested, contrary to potential models,
that $J/\psi$ and $\eta_c$ survive till temperatures as high as $1.6T_c$ 
\cite{umeda02,asakawa04,datta04}. It has been also found that $\chi_c$ 
melts at temperature of about  $1.1T_c$ \cite{datta04}. 
It is expected that there is a hierarchy of dissolution temperatures
for different quarkonia, states corresponding to hierarchy of their
sizes, i.e. we expect a sequential dissolution pattern where larger
states dissolve at smaller temperatures.
In this contribution we present our preliminary study of bottomonia
spectral functions and correlators at finite temperature.

\section{Lattice correlators and spectral functions}%

In our lattice investigation we calculate correlators of point
meson operators of the form $\bar q \Gamma q$, where $\Gamma$ 
is one of the Dirac matrices which fixes the quantum number of the
channel. We consider pseudo-scalar, vector, scalar, and axial-vector
channels which correspond to $\eta_b$, $\Upsilon$, $\chi_{b0}$ 
and $\chi_{b1}$ respectively.
The Euclidean time correlator $G(\tau,T)$ we calculate on the
lattice is an analytic continuation of the real time correlator,
$G(\tau,T)=D^{>}(-i\tau,T)$. Due to this relation the meson spectral
function, defined as
\begin{equation}
\sigma(\omega,T)=\frac{D^{>}(\omega,T)-D^{<}(\omega,T)}{2 \pi},
\end{equation} 
is related to the Euclidean correlator through the integral
representation
\begin{eqnarray}
&
\displaystyle
G(\tau,T)=\int d\omega \sigma(\omega,T) K(\tau,\omega,T)\\
&
\displaystyle
K(\tau,\omega,T)= 
\frac{\cosh(\omega (\tau-1/(2 T))}{\sinh(\omega/(2T))}.
\label{spectral}
\end{eqnarray}
Thus to reconstruct the spectral function from the lattice correlator 
$G(\tau,T)$ this integral representation should be inverted. 
Since the number of data points is less than number of degrees
of freedom (which is ${\cal O}(100)$ for reasonable discretization of
the integral ) spectral functions can be reconstructed only using the
Maximum Entropy Method (MEM) \cite{asakawa01}. In order to have sufficient
number of data points either very fine isotropic lattices \cite{datta04}
or anisotropic lattices \cite{umeda02,asakawa04} should be used.
Another difficulty arises due to the large quark mass,  
discretization errors ${\cal O}(a m_{c,b})$ are present in the
heavy quark system. To remove these discretization errors we use
the Fermilab approach \cite{fermilab,chen01,manke}.

We performed calculation of bottomonia correlators in
quenched QCD using anisotropic lattices and Wilson gauge
action. We used three value of the gauge coupling
$\beta=5.9,~6.1,~6.3$ and anisotropy $\xi=a_s/a_t=4$.
When the lattice spacing is set using the Sommer scale
$r_0=0.5$ fm the above gauge coupling correspond to
lattice spacing $a_t^{-1}=5.88,~8.18,~10.89$ GeV. 
Further details of the lattice action together with the parameters 
can be found in Ref. \cite{manke}.

To be able to reconstruct the spectral functions
high statistical accuracy for bottomonia correlators 
is required. This makes the  
calculations computationally intensive, 
so we performed them on a prototypes of RBRC QCDOC machine.
It is a dedicated Lattice QCD machine developed by physicists from Columbia
University, BNL, RIKEN and UKQCD. Three such machines, each reaching about
10TFlops peak performance, are currently installed at BNL
and EPCC. Our prototypes were single-motherboard
machine at about 50 GFlops peak. Such resources are still not adequate for the
full QCD simulations, so we use quenched approximation, which is equivalent to
neglecting quark loops. Typical statistics gathered was 500 to 1000
measurements, separated by 400 updates.

\section{Spectral functions at zero temperature}

\begin{figure}
\includegraphics[width=3in]{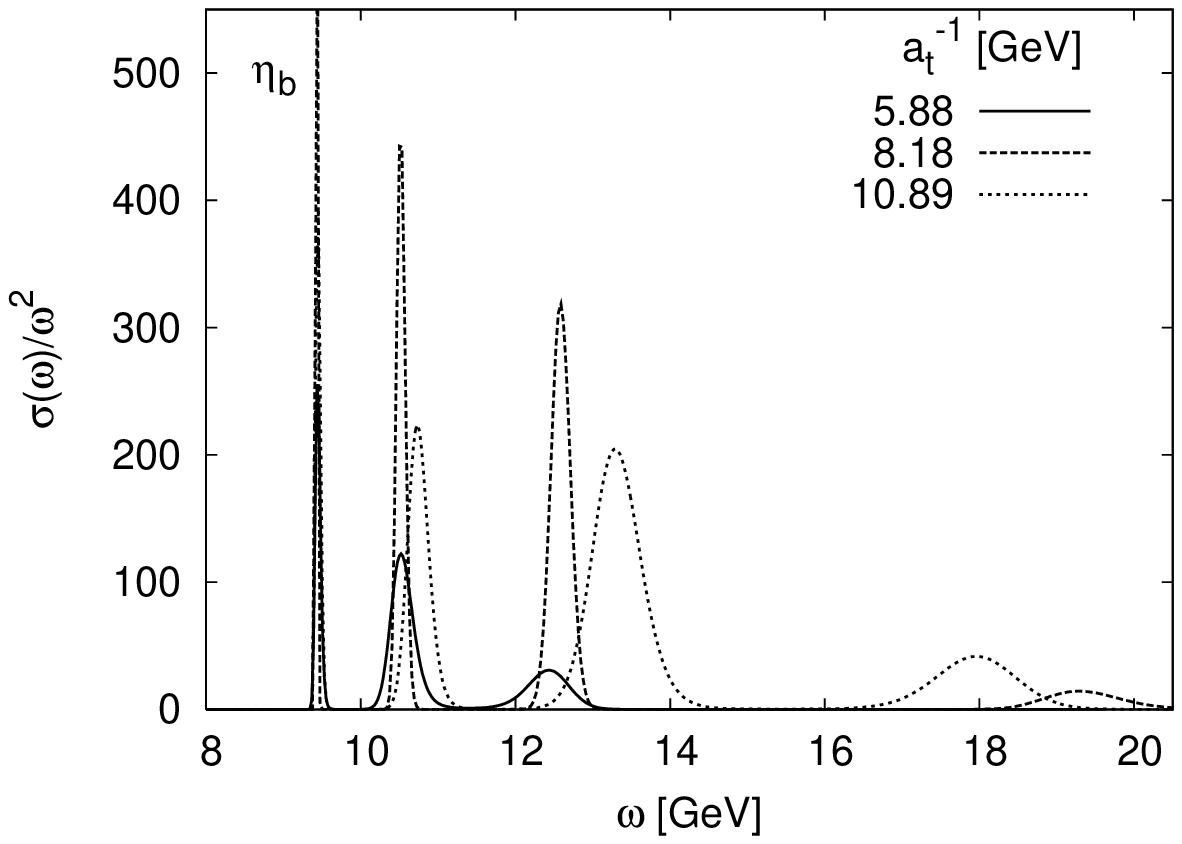}
\includegraphics[width=3in]{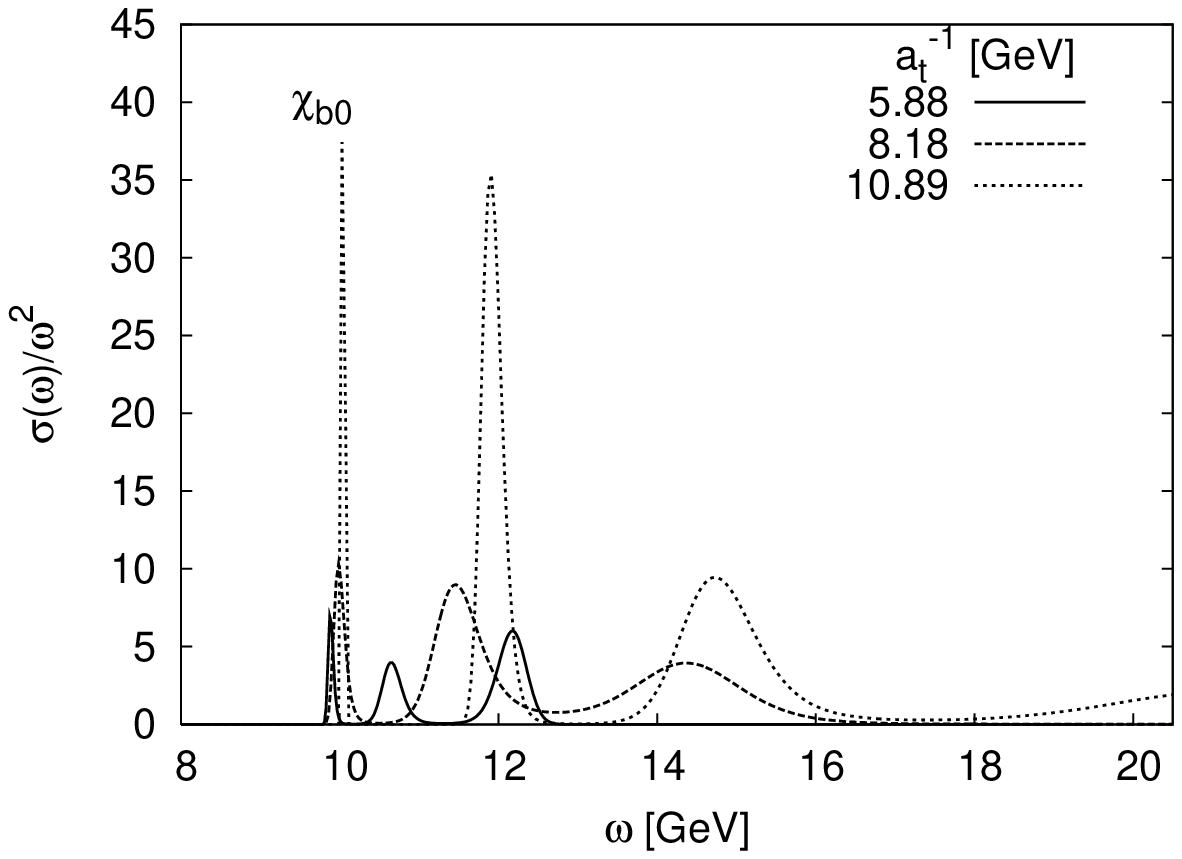}
\caption{Bottomonia spectral functions in pseudo-scalar
(left) and scalar (right) channel for different lattice
spacing.}
\label{spfzero}
\end{figure}

Bottomonia spectral functions at $T=0$ are shown in Fig.~\ref{spfzero}
for the pseudo-scalar and scalar channels.
The first peak is independent of the lattice spacing and
corresponds to $\eta_b$ and $\chi_b$ respectively.
The position of the higher peaks depends on the lattice 
spacing, so they cannot be interpreted as physical states.
Similar results have been found for the vector and axial-vector
channels.

\section{Correlators and spectral functions above deconfinement}

\begin{figure}
\includegraphics[width=3in]{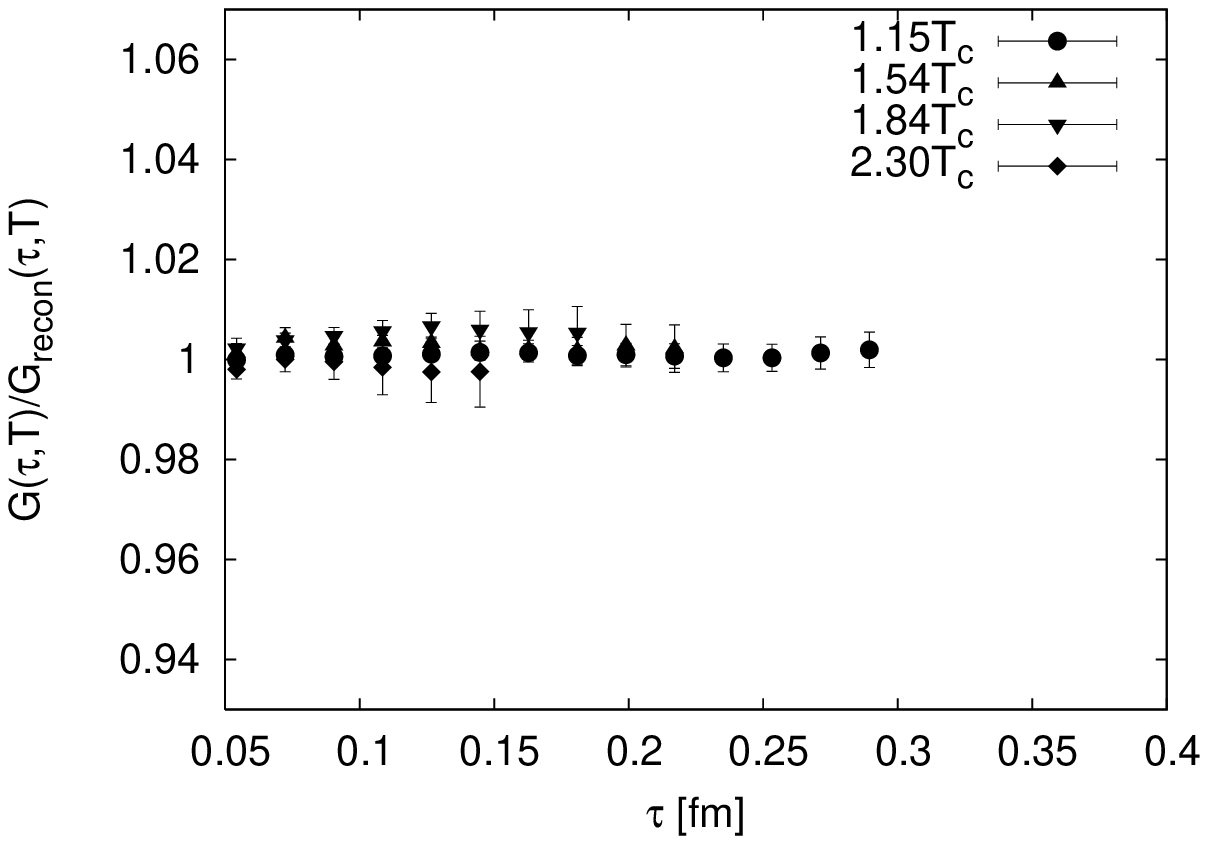}
\includegraphics[width=3in]{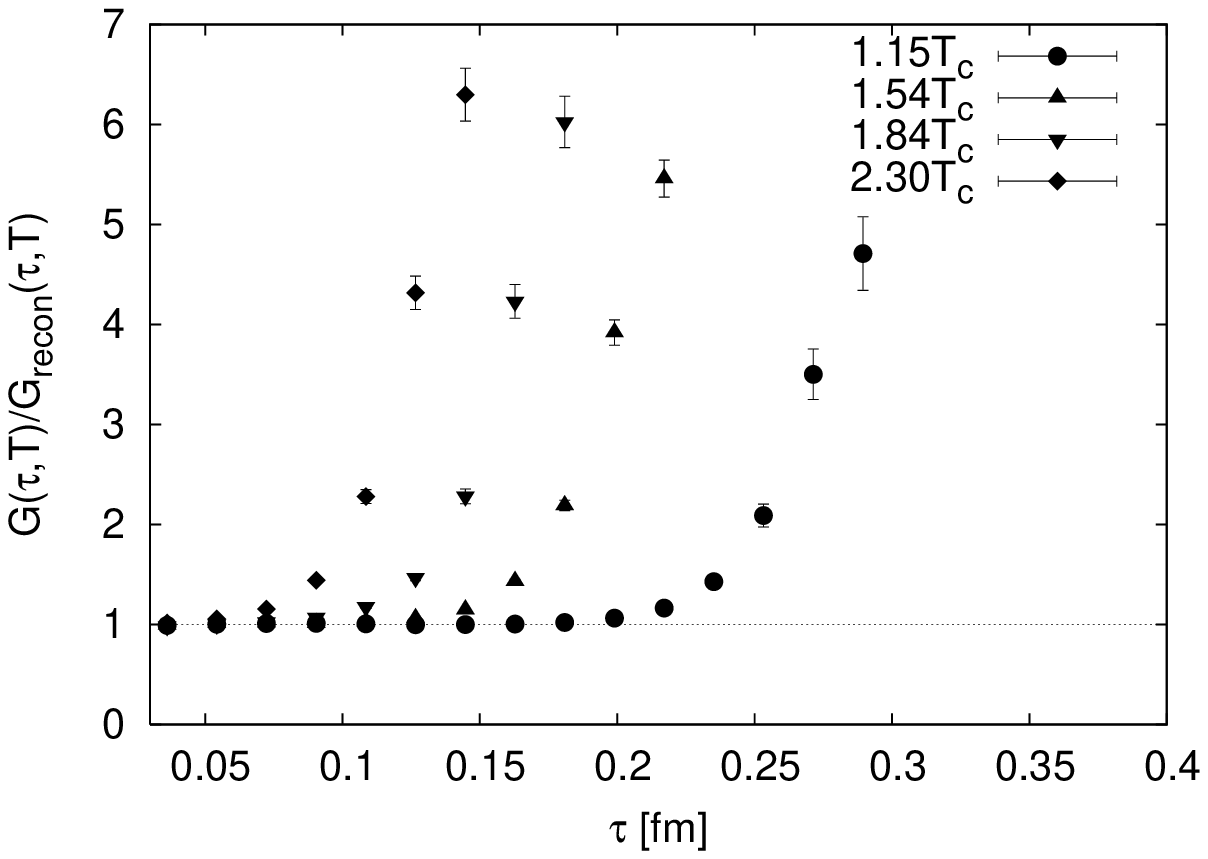}
\caption{The ratio $G(\tau,T)/G_{recon}(\tau,T)$ for different
temperatures at $\beta=6.3$.}
\label{ratio} 
\end{figure}

We would like to know what happens to different bottomonia states
at temperatures above the deconfinement temperature $T_c$. With
increasing temperature it becomes more and more difficult to
reconstruct the spectral functions as both the number of available
data points as well as the physical extent of the time direction
(which is $1/T$) decreases. Therefore it is useful to study the
temperature dependence of bottomonia correlators first. From
Eq. (\ref{spectral}) it is clear that the temperature dependence 
of bottomonia correlators come from two sources: the temperature
dependence of the spectral function and temperature dependence of
the integration kernel $K(\tau,\omega,T)$. To separate out the 
trivial temperature dependence due to the integration kernel,
following Ref \cite{datta04} at each temperature we calculate
the so-called reconstructed correlator  
\begin{equation}
G_{recon}(\tau,T)=\int_{0}^{\infty}d\omega
\sigma(\omega,T=0)K(\tau,\omega,T)
\end{equation}
Now if we assume that there is no temperature dependence 
in the spectral function - then the ratio of the original and 
the reconstructed correlator should be close to one,
$G(\tau,T)/G_{recon}(\tau,T) \sim 1$. 
This way we can identify the cases when spectral function itself 
changes dramatically with temperature. 
This gives reliable information about the fate of quarkonia states above
deconfinement. 
In Fig.~\ref{ratio} we show this ratio at $\beta=6.3$ 
for several temperatures corresponding to $N_t=32,~24,~20,~16$ for
pseudo-scalar and scalar channels. 
As one can see from the figures for pseudo-scalar channel
this ratio is close to one and shows no temperature dependence
up to temperatures $T=2.3T_c$. Thus we expect that $\eta_b$
will exist in the plasma phase with almost no modification
of its properties. On contrary in the scalar channel the correlator
shows large temperature dependence and $G(\tau,T)/G_{recon}(\tau,T)$
is far from unity already at $1.15T_c$. 
Similar behavior of the correlators was found in the axial-vector
channel. This suggest modification
or eventually dissolution of $\chi_b$ above deconfinement.
We find very similar temperature dependence of the correlators 
at larger lattice spacings corresponding to $\beta=5.9$ and $6.1$. 
This suggest that despite possible lattice
artifacts in the correlators and spectral functions medium 
modifications of bottomonia properties at finite temperature
can be reliably understood.

\begin{figure}
\includegraphics[width=3in]{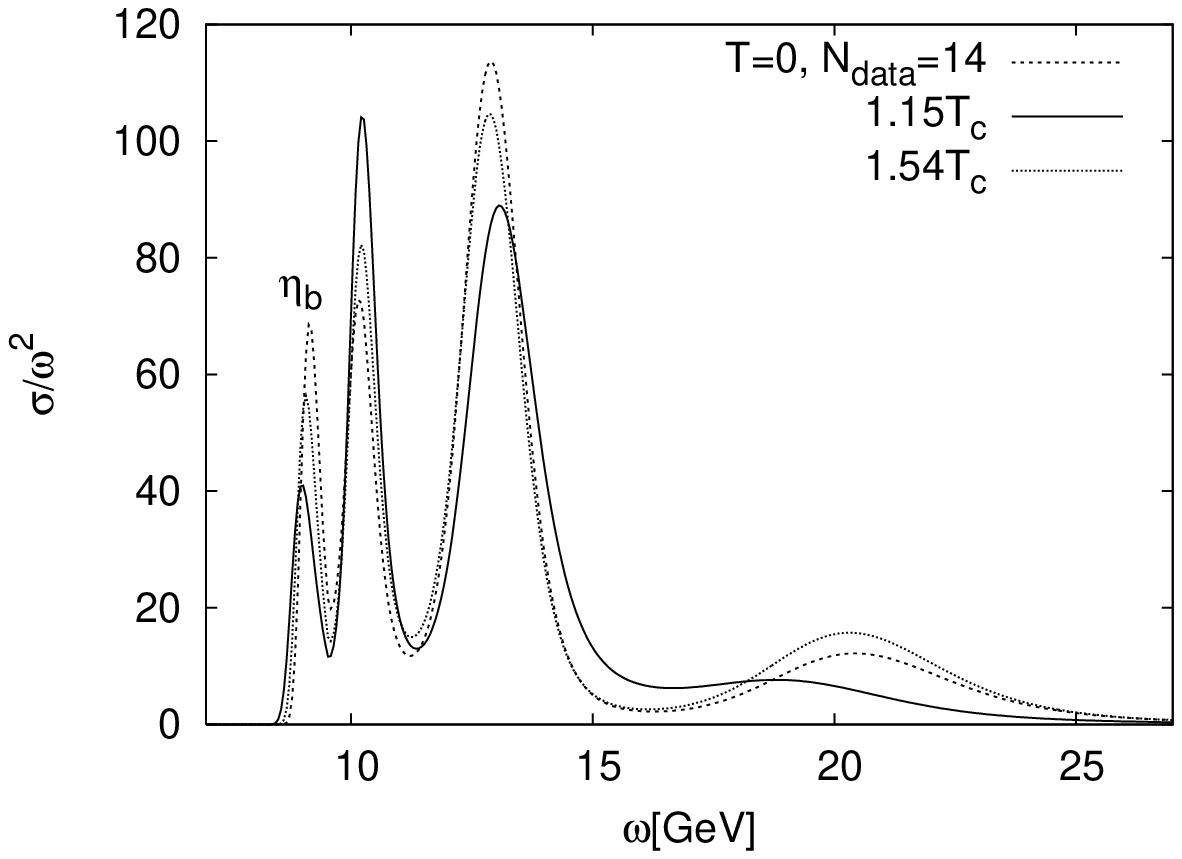}
\includegraphics[width=3in]{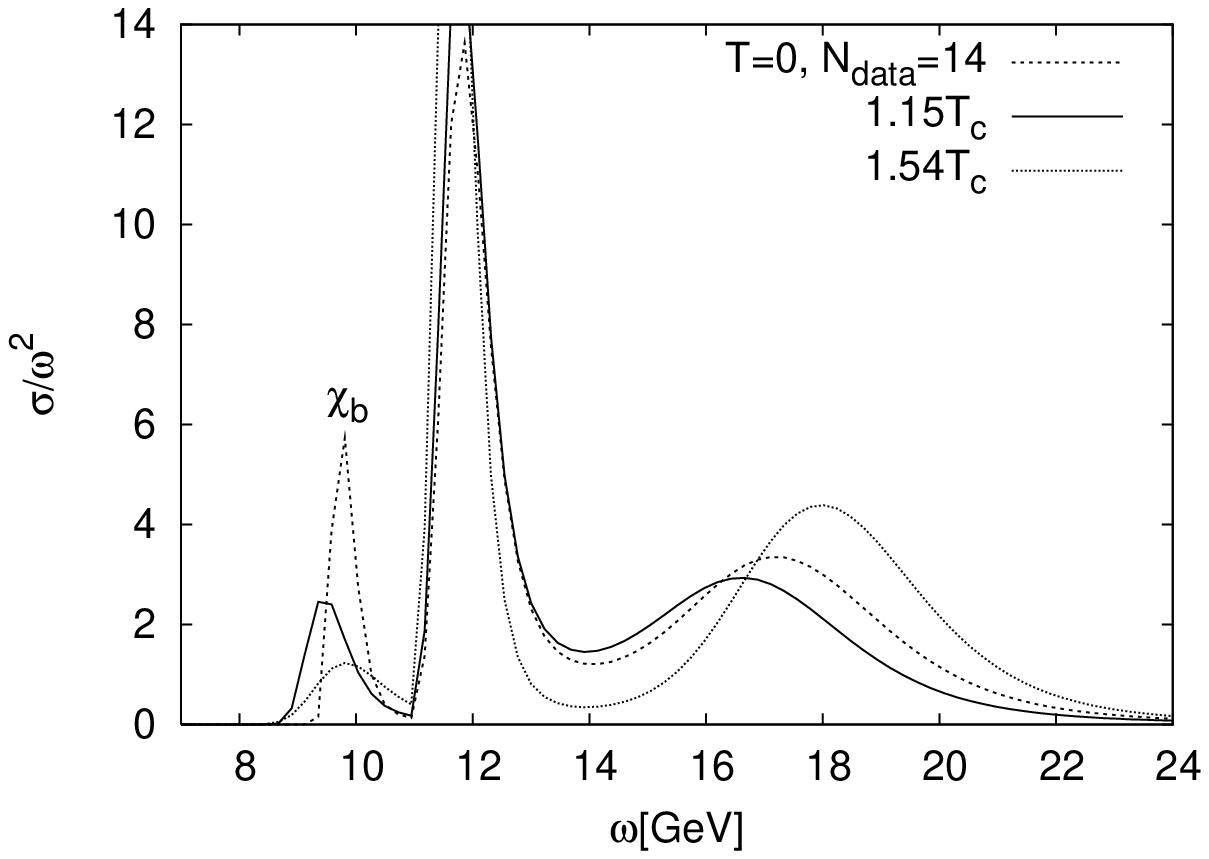}
\caption{Bottomonia spectral functions in the 
pseudo-scalar (left) and scalar channel (right)
above deconfinement for $\beta=6.3$.}
\label{figspf}
\end{figure}

More detailed information on different bottomonia
states at finite temperature can be obtained by
calculating spectral functions using MEM.
The result of these calculation is show in
Fig.~\ref{figspf}. As expected the $\eta_b$ state
survives in the deconfined phase with no mass shift.
In the scalar channel the first peak becomes broader
and smaller with increasing the temperatures, signaling
dissolution of the $\chi_b$ state. This behavior of 
$\chi_b$ state is quite unexpected as it has similar
size and binding energy as the $\eta_c$ and $J/\psi$ 
states. Thus in this respect, the expected sequential
dissolution pattern of quarkonia states does not seem
to hold.

\section{Acknowledgments} 
This work was supported by U.S. Department of Energy under 
Contract No. DE-AC02-98CH10886 and by SciDAC project. 
A.V. was partially supported by NSF-PHY-0309362.
Authors used Columbia Physics System (CPS) with high-performance 
clover inverter written by P.~Boyle and other parts 
written by RBC collaboration. 
Special thanks to C.~Jung for his generous help with CPS.

\end{document}